# Weighted Label Propagation Algorithm based on Local Edge Betweenness

Hamid Shahrivari Joghan, Alireza Bagheri, Meysam Azad

*Department of Computer Engineering and Information Technology, Amir Kabir University of Technology, Tehran, Iran*

hamid.sh.j@aut.ac.ir, ar_bagheri@aut.ac.ir, MeysamAzad@aut.ac.ir

Abstract:

In complex networks, especially social networks, networks could be divided into disjoint partitions that the ratio between the number of internal edges (the edges between the vertices within same partition) to the number of outer edges (edges between two vertices of different partitions) is high. Generally, these partitions are called communities. Detecting these communities helps data scientists to extract meaningful information from graphs and analyze them. In the last decades, various algorithms have been proposed to detect communities in graphs, and each one has examined this issue from a different perspective. However, most of these algorithms have a significant time complexity and costly calculations that make them unsuitable to detect communities in large graphs with millions of edges and nodes. In this paper, we have tried to improve Label Propagation Algorithm by using edge betweenness metric, so that it is able to identify distinct communities in both real world and artificial networks in near linear time complexity with acceptable accuracy. Also, the proposed algorithm could detect communities in weighted graphs. Empirical experiments show that the accuracy and speed of the proposed algorithm are acceptable; additionally, the proposed algorithm is scalable.

## Introduction

Generally, most of the real world problems can be modeled as complex networks. The most important challenge after modeling a problem as a graph is how to extract data and interpret it in an understandable manner. In order to analysis graphs, one must recognize the structure of the network, and communities are one of most popular structures especially in social network. Identifying community structures is an important issue in many scientific fields. After detecting communities in a graph, sub-sections are obtained in which its members have a strong connection to each other while connections between sub-sections are minimized. Detecting communities is related to other concepts like social networks analysis (relationships among members), biological research, or technological issues (optimizing large infrastructure).

As mentioned before, the interaction between nodes within a community is more than nodes outside the community. For instance, when we detect communities in friendship networks, the output partitions are real world friendship groups. At the network level, the vertices may be divided into several different categories, while each group forms a community. The number of different states for categorizing vertices is exponential [1]. According to the research done, the complexity of detecting the best categorization, connectivity in communities is maximized while the connectivity between communities is minimized, is NP-complete [2].

In this paper, we introduce Weighted Label Propagation Algorithm based on Local Edge Betweenness, abbreviated as WLPA-LEB that is a random based algorithm using edge betweenness metric as heuristic to increase accuracy of traditional LPA. The proposed algorithm is capable of identifying communities in weighted networks. We used local edge betweenness in order to keep time complexity of the whole algorithm near linear. During detecting communities, our algorithm propagates label more efficiently by taking label from edges with the lower local edge betweenness. In fact, edges in a same community usually have small edge betweenness value while edges between the two communities have large value [3]. As a result, our algorithm gives less chance to nodes to get a label from nodes outside its real community.

The results presented by our algorithm and other algorithms show that the accuracy of our algorithm is acceptable. Experiments have been made on a various real-world directed and weighted networks; besides, the efficiency of the algorithm in the artificial networks has also been verified with satisfactory results. In addition to the above, the scalability and time complexity of the algorithm have been investigated in practice, which indicates the algorithm's high performance.

The rest of the paper is organized as follows: in the preliminary section, we describe graph definition and notation; also we detail LPA and local edge betweenness. In the related work section, we review previous works and algorithms, most of which are based on random approaches. In the purposed algorithm section, we define WLPA-LEB, and detail each step of the algorithm. In the experiments section, both experiments on real-world networks and artificial networks are compared. We also examine scalability of the proposed algorithm on shared memory model. Finally, conclusion and future works are presented at the end of the paper.

## Preliminary

A complex network consists of several vertices connected by the edges. A vertex can represent a person, user, geographic coordinates, city, processing unit, or any identifiable entity that connects to other entities if they have a relationship. Formally, graph ($G$) is represented as $G= (V, E)$ that $V$ stands for nodes and $E$ for edges, also $n$ is the number nodes and $m$ is the number of edges. Let $d_i$ denote the degree of the node $i$ and $x_i$ denote label of the node $i$, also let $d_{ii}(C)$ denotes the number of edges of node $i$ which are in $C$ community, and $d_{iout}(C)$ denotes the number of edges of node $i$ which are not in $C$ community. If the edges have direction, each node has the degree in $d_i^i$ and degree out $d_i^{out}$, and $d_i^i$ is equal to the number of edges which their destination is node $i$, $d_i^{out}$ is the number of edges which their source is node $i$.

### Community

The researchers have agreed on a simple network graph model that is called the l-partition model [4]. Each partition in this model contains a number of vertices that are connected to the vertices in the same community with the probability $P_i$, and is connected to vertices outside its community with the probability $P_{out}$. As long as $P_i \geq P_{out}$, the graph contains communities, otherwise, the network is only a random graph that does not have a specific community structure. The most popular model of l-partition is the GN model [5]. In this model, the network graph contains 128 vertices with an expected degree of 16, which is divided into 4 groups of 32. This model is used to compare the efficiency of community detection algorithms; what was done in 2005 by Danone et al. [6] on the most famous algorithms of that time. However, this model has two major drawbacks: First, all the vertices have an equal degree; Second, all communities are the same size. These disadvantages have been resolved in the LFR mode [7]. This model is much more rigorous and limits algorithms faster, while in the GN model it seemed that the algorithms had a good performance, but the reason for this successful experiment was the simplicity of GN model. In addition, the LFR model is constructed in linear time; therefore, it is also suitable for large graph testing.

In another model, the community is a sub graph of the network that the number of internal edges between the vertices of the sub graph is higher than the external edges of the sub graph [8]. To increase the accuracy of the algorithms, we need to provide a more precise definition for the community. Reasonable and varied definitions are given in different articles; however, we use the definition purposed in [8] named as strong and weak communities. The graph has strong community if for each node $i$ in community $C$:

$$d_{in}(C) > d_{iout}(C)$$

Also, we can define weak community while each community $C$ obtains below equation:

$$\sum_{i \in C} d_{in}(C) > \sum_{i \in C} d_{iout}(C)$$

In a weak community, the sum of edges number in the community is bigger than the sum of edges number out of the community.

**LPA**

In 2007, Raghvan et al [9] proposed a random algorithm that can quickly identify communities in complex networks. In this approach, all the vertices have a unique label, and at each iteration, each vertex selects the label that has the most frequency in its neighbors and replaces it as own label. This procedure continues until all the vertices of the graph have obtained the most frequent label of its neighbor. Eventually, the nodes sharing the same label are in the same community. The following steps outline the algorithm in detail:

1. Each vertex gets a unique label.

2. The vertices are randomly stored in list *X*

3. For each *i*∈*X* chosen in that specific order, *i* get the most frequent label of its neighbors.

4. If all the vertices do not have the most frequent label, go to step 2.

5. Nodes with the same label are in the same community.

The complexity of the first, second, and fifth steps is $O(n)$, but the complexity of the third and fourth steps is $O(m)$ because each vertex must check its neighbors. Since the order of $O(n)$ is less than $O(m)$, the complexity of the entire algorithm is equal to $O(m)$. On the other hand, the number of iterations in the fourth step is usually constant and, according to the author, in 95% of the cases, algorithm finds the right answer in less than 5 rounds.

**Local Edge-Betweenness**

Edge-betweenness [10] is a measure of edge importance which indicates the number of shortest paths between pairs of vertices that paths through it. In the case that more than one shortest path exists between a pair of vertices, each path obtains an equal weight. Newman and Girvan (GN) [10] use edge-betweenness in order to remove edges between communities, but the time complexity of the algorithm is $O(n \cdot m^2)$; therefore, in large graphs, calculating edge-betweenness is inefficient.

In another article [11], Steve Gregory introduces Local Betweenness in order to reduce the complexity of Girvan-Newman algorithm. In order to calculate local edge-betweenness of edge *e*, instead of counting all shortest paths, just counts *h* depth shortest paths running along *e*. WLPA-LEB uses 2-depth local edge-betweenness to distinct between edges which are in the same community and those are not. The time complexity of calculating edge-betweenness for all edges is $O(n.m)$, because it adapts BFS on each node to find all shortest paths between pairs; however, h-depth local edge-betweenness requires h-depth BFS, so 2-depth local edge-betweenness complexity is $O(n.(\frac{m}{n})^2)$ and in sparse graph is $O(n)$ which is linear-time.

## Related works

In paper [12] LPAm was proposed which examined the label propagation algorithm as an optimization problem. However, it was prone to get stuck in poor local maxima in the modularity space. Therefore, a multistep greedy agglomerative algorithm (MSG) is proposed that can merge multiple pairs of communities at a time. After that, an advanced modularity-specialized label propagation (LPAm+) is proposed [13]. In another paper [14], Kang and Jia proposed an improved version of LPA called label propagation algorithm based on local similarity (LPALS) in which the similarity was used to be the weight value of the node labels.

Y. Xing et al. [15] proposed NIBLPA, which updates node's label based on fixed node orders of labels updated in the descending order of node importance. In another paper, R. Francisquini [16] tries to use a meta-heuristic algorithm and introduces GA-LP, that its main idea is the local nature of the key operators of the genetic algorithm. X. Zhang et al. [17] represent LPA-NI which is label propagation algorithm for detecting communities based on node importance and label influence. Zhao et al [18] proposed a novel algorithm for community detection called Label-Influence-Based (LIB) which selects a set of nodes as the seeds and the label propagation procedure begins from the seeds. H. Lou et al [19]presented weighted coherent neighborhood propinquity (weighted-CNP) to calculate the probability that a pair of vertices is involved in the same community, in order to improve LPA accuracy.

Based on the mutual information of direct and indirect neighbors, N. Chen et al. [20] introduced LPA-E which is capable of detecting both overlap and disjoint community. X. Zhang et al. [21] presented LPALC which is also able to detect overlapping communities. LPALC is based on LPA and improves the propagation process by choosing the nearest neighbor having a local cycle instead of choosing a neighbor randomly. J. Xie et al. [22] proposed SLPA that unlike traditional LPA, holds more than one label for each node, so it is capable of detecting overlapping communities, too. Gregory puts forward COPRA [23] which can detect overlapping communities. Same as SLPA strategy, COPRA assigns multiple labels to each node.

In another work, X. Zhang et al [24] introduced a random-based algorithm (LPAc) based on edge clustering coefficient [8]. LPAc strategy is that node's label whose edge clustering coefficient is higher to be propagated preferentially. It is worthy to mention, LPAc is the most similar algorithm to WLPA-LEB. There are different algorithms to detect communities, but this section just focuses on LPA based algorithms. For more details, many surveys compare variant approaches and algorithms for detecting both disjoint and overlapping communities [25] [26] [27] [28] [29].

## Proposed Algorithm

In this section, we introduce the proposed algorithm, Weighted Label Propagation Algorithm based on Local Edge-Betweenness which is briefly called WLPA-LEB. As noted, the accuracy of the LPA algorithm is low, but its complexity in sparse networks is linear; on the other hand, the Girvan-Newman algorithm has a high accuracy but high complexity. We can merge these two algorithms and propose a fast algorithm with acceptable accuracy. In this proposed algorithm, we have tried to improve the accuracy of the random technique introduced in the LPA by means of edge-betweenness metric. Besides, WLPA-LEB has also been improved to detect community in weighted graphs, but it does not hold specific approach to detect community in directed networks; indeed, we assume all directed networks as undirected ones.

One of the factors for low accuracy in the LPA is that there is no difference between neighbors of a node. In fact, a chance of obtaining a label from a neighbor in the same community is equal to a neighbor outside the community. Especially, at the initial iterations, when communities are small and weak, if we give a higher chance to neighbors in the same community, then the algorithm convex faster and more accurate. As mentioned before, edges between communities have a high edge betweenness value while the edges inside the same community have lower value; thus, we can give higher chances to neighbors with the low edge betweenness since with high probability, these neighbors are in the same community. Yet, two important points should be considered:

1- The complexity of computing the edge betweenness is $O(n.m)$ which is higher than the complexity of the LPA.
2- How to use edge betweenness metric in order to increase the chance of getting labels from neighbors in the same community.

To answer the first problem, we tried to use local edge betweenness because communities are a local concept and local metrics can also identify communities with high precision. Also, local betweenness is presented in a research paper [11] in the form of an improved version of the Girvan-Newman algorithm that represented acceptable results. In local edge betweenness, all the shortest paths within the network are not investigated; instead, other short paths, such as two or three depths (*h*), are examined. As mentioned before, the complexity of computing *h* depths local edge betweenness is $O(n.(\frac{m}{n})^h)$.

By calculating 2-depth local edge-betweenness of a Zachary Club Friendship Network [30] in Figure 1, it is clear that the edges between communities have bigger value than edges inside the same community, so the possibility of obtaining label from neighbors with high values should be decreased. At this point of our algorithm, we assume that all edges have the same weight, but we affect the weights of the edges in the next steps.

In order to answer the second problem, WLPA-LEB has an extra step in which nodes obtain label from just half of its neighbors. Based on traditional LPA, if half of a node's neighbors have the same label *i*, that node belong to community *i*; on the other hand, neighbors with low local edge-betweenness are more likely in the same community. Thus, at first, WLPA-LEB sorts each node's neighbor based on their local edge betweenness; secondly, each node obtains the label from half of its neighbors which they have lowest local edge betweenness. In this step, WLPA-LEB actually gives a higher chance to neighbors with the lower local edge betweenness, but other neighbors get zero probability since just half of the neighbors have been examined. In the next step, same as the traditional LPA, each node obtains the label from all of its neighbors, so all neighbors have non-zero probability.

The first step tries to merge high potential labels since labels come from neighbors with low local edge betweenness; in fact, cores of communities are emerging. The second step, which has no limitation, tries to extend small and strong communities to bigger ones based on traditional LPA criteria. In both steps, if a node has more than one most frequent label, it randomly chooses one of them. Indeed, all neighbors have non-zero probability while high potential labels have a higher chance to be propagated. Steps of WLPA-LEB are detailed as follows:

1. Each vertex gets a unique label
2. Calculate h-depth local edge betweenness
3. Sort each node's neighbors based on their local edge-betweenness value.
4. The vertices are randomly listed in the list $X$
5. For each $i \in X$ chosen in that specific order, node $i$ obtains the most frequent label (label edges weight for weighted networks) from half of its neighbors which they have lowest local edge betweenness.
6. The vertices are randomly listed in the list $X$
7. For each $i \in X$ chosen in that specific order, node $i$ gets the most frequent label of its neighbors.
8. If all the vertices do not have the most frequent label, go to step 4.
9. Nodes with the same label are in the same community.

For weighted networks, we rewrite the fifth step that instead of half of neighbors' count, it obtains the label from neighbors which they have lowest local edge betweenness and their edges weight aggregation is not bigger than half of the node's total edges weight. In fact, in a weighted graph, a vertex obtains label $i$ if the edges weight having label $i$, is greater than other labels' edges weight. Therefore, in weighted graphs, the criterion is not the number of edges, but the weight of the edges is the criterion.

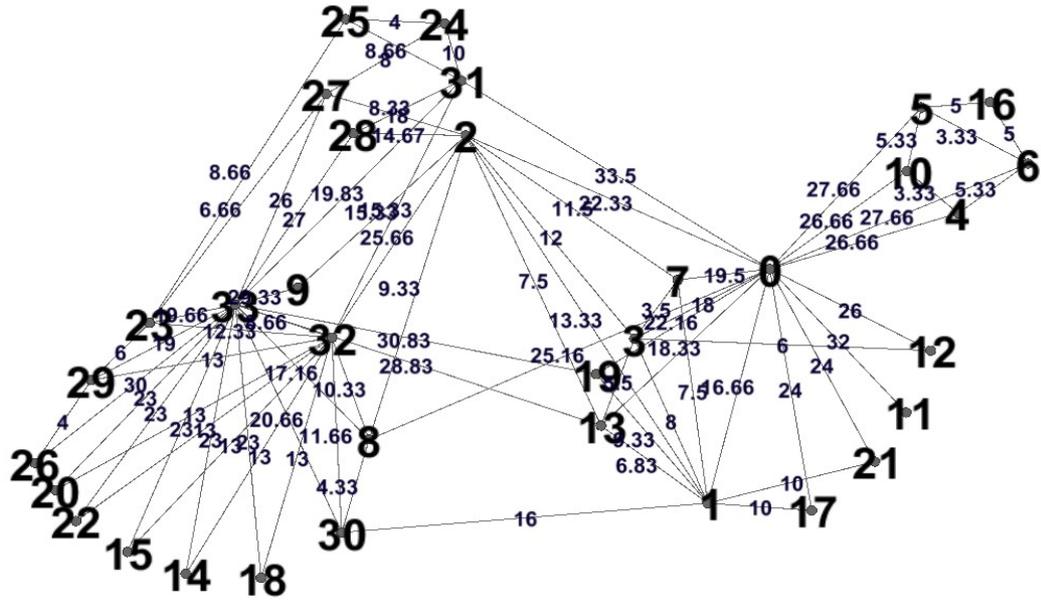

Figure 1, Zachary karate club network that edges value is 2-depth local edge-betweenness. Edges between communities have higher local edge-betweenness compared edges inside communities.

We describe the procedure of the algorithm over a real-world network named Zachary karate club. Figure 1 represents the graph which each vertex has a unique label; also, the value of edges indicates 2-depth local edge-betweenness. Nodes' label, after the first iteration of step 5 and 7, is shown in Figure 2. Highly to mention, at each run, the answers will be different because the algorithm is random. As is pellucid, communities have received almost the same labels, but the algorithm will not end because the step 8 condition has not yet been met. Figure 3 represents nodes label after the second iteration. Each community has its own unique label, so the algorithm ends. Yet, two vertices (*a*) and (*b*), as defined in the figure 3, do not have the correct label which is 33, but they still satisfy the stop criteria, and the algorithm terminates. However, the main idea is that, at each run, WLPA-LEB does not necessarily find the best answer, but if we run the algorithm several times, the correct answer will be among them. In another run, these two vertices may randomly obtain correct label which leads to a better answer.

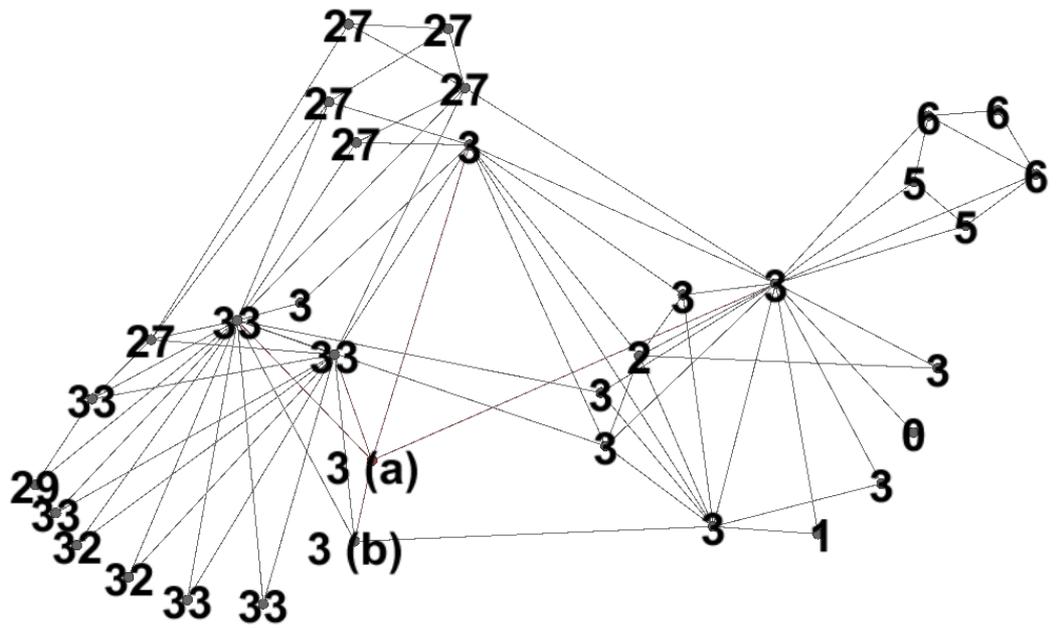

Figure 2, nodes label after WLPA-LEB first iteration. Nodes (*a*) and (*b*) did not obtain the best label *33*.

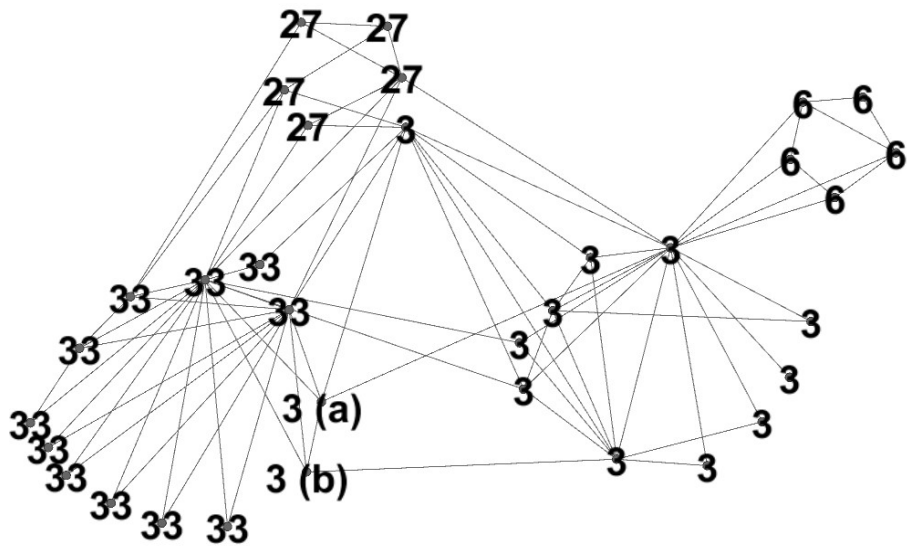

Figure 3, nodes label after WLPA-LEB second iteration. Detected communities satisfy step 8 stop criteria, and the algorithm terminates. Nodes (*a*) and (*b*) did not obtain the best label *33*, but the algorithm is random and it would find better answer in other runs.

To calculate the complexity of proposed algorithm, the complexity of each step must be calculated. Steps 1, 4, 6, and 9 are order of $O(n)$, and the order of steps 5, 7, and 9 is $O(m)$. The order of step 3 is $O(n.(\frac{m}{n})^2)$ because each vertex needs to sort its neighbors based on their local edge-betweenness value, and the order of step 3 is also $O(n.(\frac{m}{n})^h)$ which depends on the depth of local edge-betweenness. On the other hand, WLPA-LEB is an iterative algorithm same as LPA, and the number of iterations (*t*) directly affects the complexity of the algorithm. Ultimately, the complexity of the algorithm is equal to:

$$O(n.\left(\frac{m}{n}\right)^h + t.m)$$

In the experiments, we find out that 2-depth local edge betweenness gives acceptable results and deeper levels cost does not worth the accuracy improvement. Also, the algorithm with less than 10 runs can detect the correct answers. Eventually, in the sparse graphs, the complexity of this algorithm is almost linear since the number of edges has a linear relationship with the number of vertices. The spatial complexity of proposed algorithm is order of $O(m)$ which is almost linear.

## Experiment

In this part of the paper, we evaluate the proposed algorithm, and its accuracy and speed are compared to other existing algorithms. During experiments, both real-world networks and synthesized networks have been used. All experiments are practical and the results are achieved under the same conditions and criteria. At the end of this section, the speed of the algorithm is practically compared and analyzed, and the scalability of this algorithm is evaluated. The Weighted Label Propagation Algorithm based on Local Edge-Betweenness has represented acceptable results in undirected weighted networks.

## Dataset

As mentioned above, experiments are both on real-world and synthesized networks. Table 1 represents real-world networks and their properties: besides, Table 2 shows synthesized networks generated by LFR [7]. This graph generator can generate various networks with different properties like node count (*n*), edge count (*m*), average degree ($\frac{m}{n}$), maximum average degree $max(\frac{m}{n})$, maximum community size (*maxc*), minimum community size (*minc*) and generates weighted or directed network.

## Evaluation Metrics

In order to compare detected communities quality in real-world networks, we use modularity [31] which assigns positive score to edges inside communities and negative score to edges between communities. Modularity ranges between -1 and +1 that positive values indicate good community structures. To evaluate algorithms performance in synthesized networks, we can use NMI [32] since LFR produces ground truth communities, and its value ranges between 0 and +1.

Table 1, real-world networks properties including name, node count *(n)*, edge count *(m)*, and average node degree ($\frac{m}{n}$)

| Name | Node count (*n*) | Edge count (*m*) | Average node degree ($\frac{m}{n}$) |
|---|---|---|---|
| Karate [30] | 34 | 78 | 458 |
| Dolphin [33] | 62 | 159 | 5.12 |
| Football [10] | 115 | 613 | 10.66 |
| Power [34] | 4941 | 6594 | 2.66 |
| Internet[1] | 22963 | 48436 | 4.21 |

## Algorithms for Comparison

Various technics and algorithms are presented that detect community structures in complex networks. For comparison and evaluation, we have selected below algorithms:

1. COPRA [23], which is a random method that has been studied in most articles and is capable of detecting weighted and directed communities.
2. Label Propagation Algorithm (LPA), which the proposed algorithm is an improved version of this algorithm. In this algorithm, we assume that all networks are undirected and unweight.
3. Girvan-Newman (GN), which is a definite algorithm and use edge betweenness to remove edges between communities.
4. LPAc [24], which is an improved version of LPA that has tried to increase the accuracy of the algorithm by means of edge clustering coefficient.

All experiments are executed on a specific system: a quad-core Core i7 processor and 16GB of internal memory. All random algorithms are executed 100 times to find the best answer.

Table 2, properties of networks generated by LFR. Features are node count (*n*), edge count (*m*), average degree ($\frac{m}{n}$), maximum average degree $max(\frac{m}{n})$, maximum community size (*maxc*), minimum community size (*minc*) and is weighted network.

| Name | n | $\frac{m}{n}$ | $max(\frac{m}{n})$ | maxc | minc | Weighted | Description |
|---|---|---|---|---|---|---|---|
| A | 1,000 | 20 | 50 | 50 | 10 | No | Small graph with small communities |
| B | 1,000 | 20 | 50 | 100 | 20 | No | Small graph with big communities |
| C | 5,000 | 20 | 50 | 50 | 10 | No | Big graph with small communities |
| D | 5,000 | 20 | 50 | 100 | 20 | No | Big graph with big communities |
| E | 100,000 | 50 | 200 | 1,000 | 20 | No | Very big graph with various |

---

1 A symmetrized snapshot of the structure of the Internet at the level of autonomous systems, reconstructed from BGP tables posted by the University of Oregon Route Views Project. This snapshot was created by Mark Newman from data for July 22, 2006 and is not previously published.

| | | | | | | | communities |
|---|---|---|---|---|---|---|---|
| F | 5,000 | 20 | 50 | 50 | 10 | Yes | Big graph with small communities |
| G | 5,000 | 20 | 50 | 100 | 20 | Yes | Big graph with big communities |

**Results**

In this part, we compare the results of WLPA-LEB with other algorithms. Experiments show that the proposed algorithm gives an acceptable accuracy versus the given cost on real-world networks.

WLPA-LEB is a random algorithm, that is, each time it gives a different answer; hence, we compare the results of the 100-times execution of the proposed algorithm, LPA, and LPAc. Results are gathered in Table 3. As can be seen, the average modularity of detected communities by the proposed algorithm are better than other algorithms that means local edge-betweenness could help labels propagate more efficiently; also, the best answer is often found by the proposed algorithm. However, for the Power network, the results of the proposed algorithm are not acceptable since the network is a tree shape.

Table 3, quality of detected communities by WLPA-LEB, LPA and LPAc in 100 runs on real-world networks. The evaluation metric is modularity. * means that the algorithm has hold the network as a single community

| Network | Average modularity | | | Best modularity | | | Worst modularity | | |
|---|---|---|---|---|---|---|---|---|---|
| | WLPA-LEB | LPA | LPAc | WLPA-LEB | LPA | LPAc | WLPA-LEB | LPA | LPAc |
| Karate | **0.3906** | 0.3455 | 0.2939 | **0.4155** | 0.4020 | 0.3717 | **0.1120** | * | * |
| Dolphin | **0.5152** | 0.4819 | 0.4897 | **0.5267** | 0.5267 | 0.5246 | **0.4350** | 0.3017 | 0.3322 |
| Football | **0.5980** | 0.5899 | 0.5791 | **0.6045** | 0.5973 | 0.6033 | **0.5330** | 0.5299 | 0.5148 |
| Power | 0.7844 | **0.8002** | 0.6860 | 0.7894 | **0.8126** | 0.6933 | **0.7787** | 0.7785 | 0.6797 |

In Table 4, the comparison between the accuracy of the proposed algorithm, LPA and COPRA is presented. As is evident, the best answer based on modularity criteria is obtained by the WLPA-LEB. In some cases, WLPA-LEB has detected communities more accurately than GN algorithm while GN computation is heavier than WLPA-LEB. In all cases, the WLPA-LEB results are better than LPA and COPRA which are random approaches.

Table 4, comparison between modularity of detected communities by WLPA-LEB, COPRA, GN and LPA. Random algorithms are executed 100 times and the best result is selected.

| **Network** | **WLPA-LEB** | **COPRA** | **GN** | **LPA** |
|---|---|---|---|---|
| Karate | 0.4155 | 0.3742 | 0.3632 | 0.402 |
| Dolphin | 0.5264 | 0.4805 | 0.5193 | 0.4824 |
| Football | 0.6045 | 0.6032 | 0.6206 | 0.5973 |
| Power | 0.7894 | 0.3192 | 0.8148 | 0.8126 |
| Internet | 0.5722 | 0.3004 | 0.5965 | 0.5647 |

As mentioned before, we have generated various synthesized networks with LFR which could generate networks with different level of community structures. In fact, LFR use a parameter named *mu* that sets the clarity of constructed communities. This parameter acts like $P_i$ and $P_{out}$ in l-partition, and ranges between 0 and 1. The value of *mu* indicates the ratio between nodes edges outside communities and edges inside communities. As *mu* grows, detecting communities become harder, and when *mu* exceeds 0.5, most of community detector algorithms could not find acceptable communities.

In our experiments on synthesized networks, we compared the NMI of detected communities with true communities represented by LFR. We have evaluated performance of WLPA-LEB, COPRA, and LPA with different values of *mu* for networks in Table 2. Results are represented in Figure 4, 5, 6 and 7.

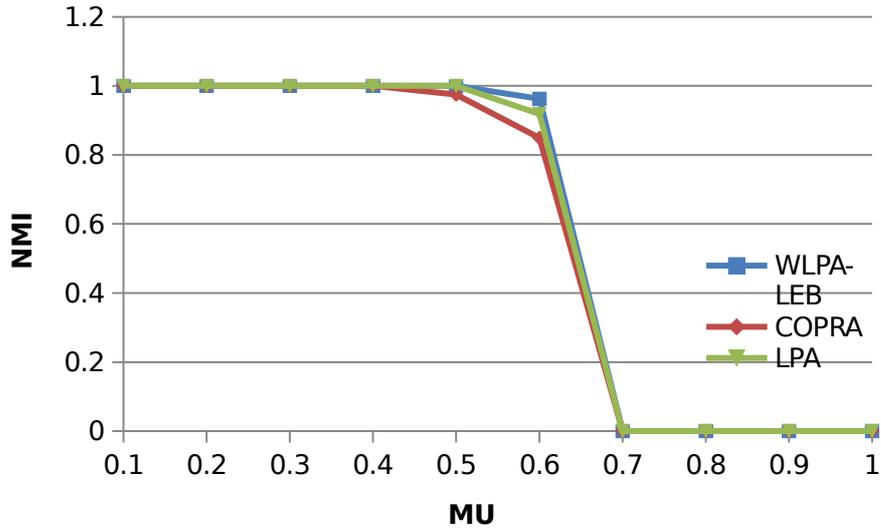

Figure 4, performance of WLPA, COPRA and LPA on small synthesized network with small communities (A). Results are compared through different value of mu. When mu exceeds 0.5, detected communities become unclear, yet WLPA-LEB could detects communities better than other algorithms

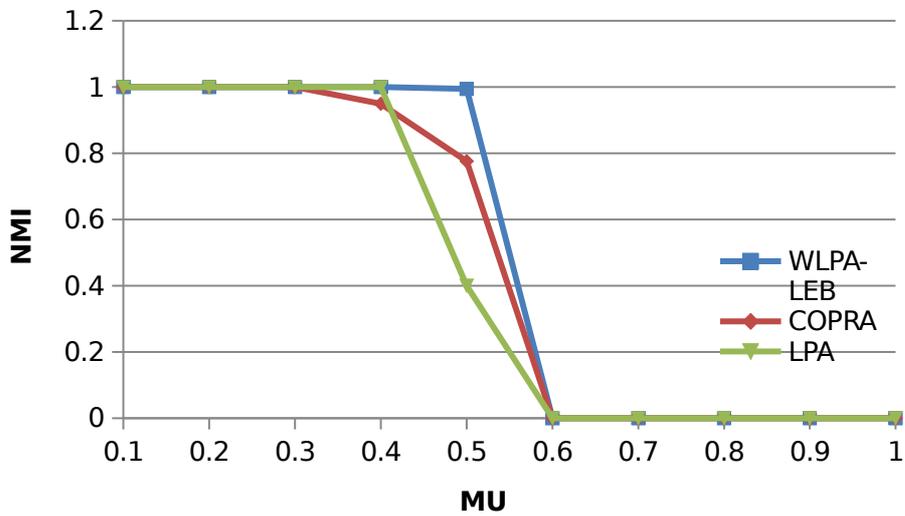

Figure 5, performance of WLPA, COPRA and LPA on small synthesized network with big communities (B). Results are compared through different value of mu. When mu exceeds 0.5, detected communities become unclear, yet WLPA-LEB could detects communities better than other algorithms

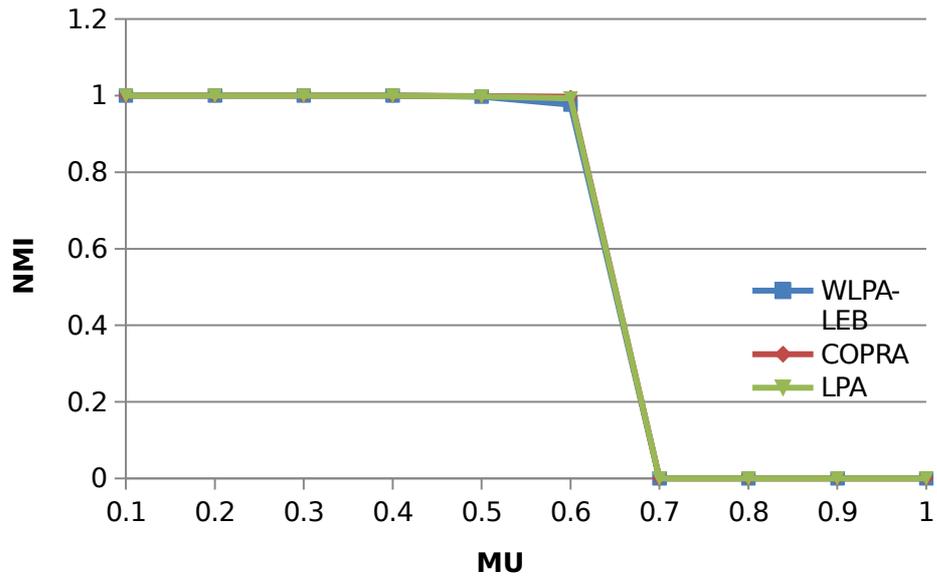

Figure 6, performance of WLPA, COPRA and LPA on big synthesized network with small communities (C). Results are compared through different value of mu. When mu exceeds 0.6, detected communities become unclear, yet WLPA-LEB could detects communities.

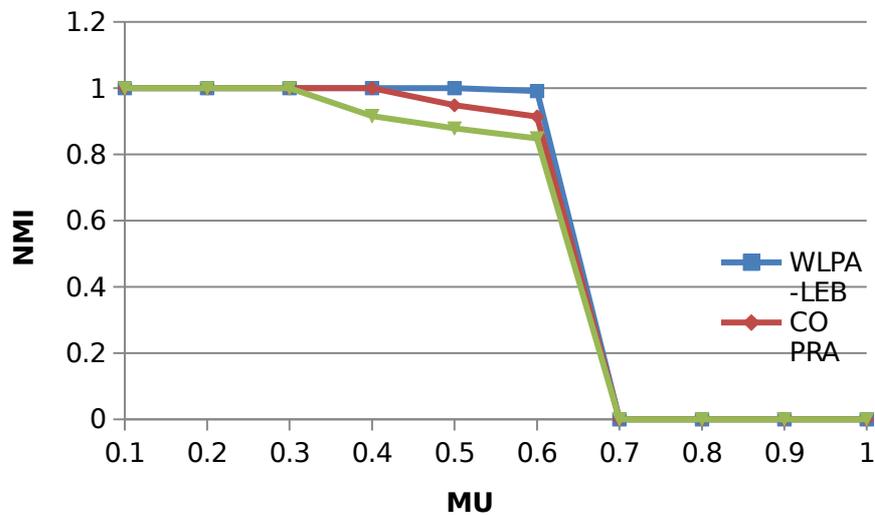

Figure 7, performance of WLPA, COPRA and LPA on big synthesized network with big communities (D). Results are compared through different value of mu. When mu exceeds 0.5, detected communities become unclear, yet WLPA-LEB could detects communities better than other algorithms.

WLPA-LEB could detect communities more efficiently when they become big because local edge-betweenness help labels propagate inside communities. Also we have evaluated performance of WLPA-LEB in a very big network with various sized communities in Figure 8. WLPA-LEB could detect various sized communities better than COPRA and LPA.

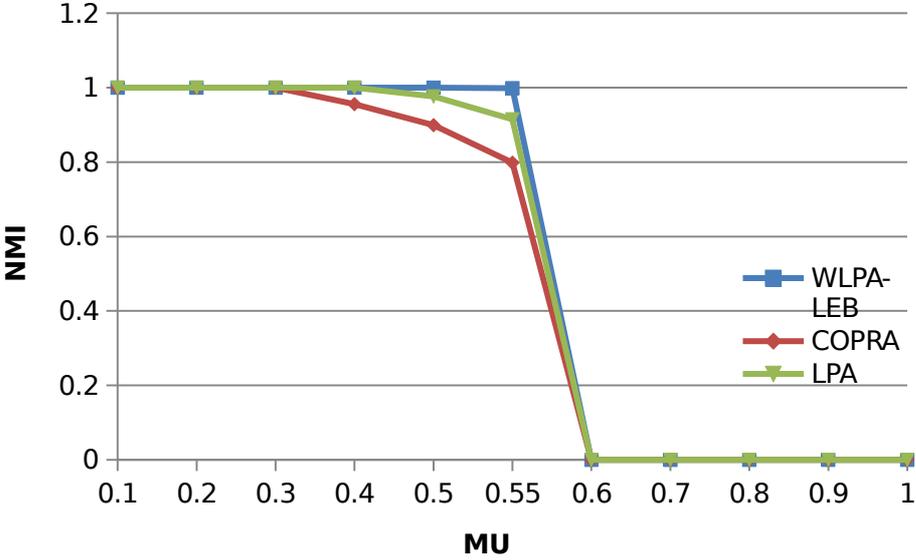

Figure 8, performance of WLPA, COPRA and LPA on very big synthesized network with variuous communities size (E). Results are compared through different value of *mu*. When *mu* exceeds 0.5, detected communities become unclear, yet WLPA-LEB could detects communities better than other algorithms.

In order to evaluate WLPA-LEB performance in weighted networks, we compared the quality of detected communities by proposed algorithm with COPRA which could detect communities in weighted networks. Results are represented in Figure 9 and 10 that indicate WLPA-LEB detects weighted communities in weighted networks with acceptable accuracy.

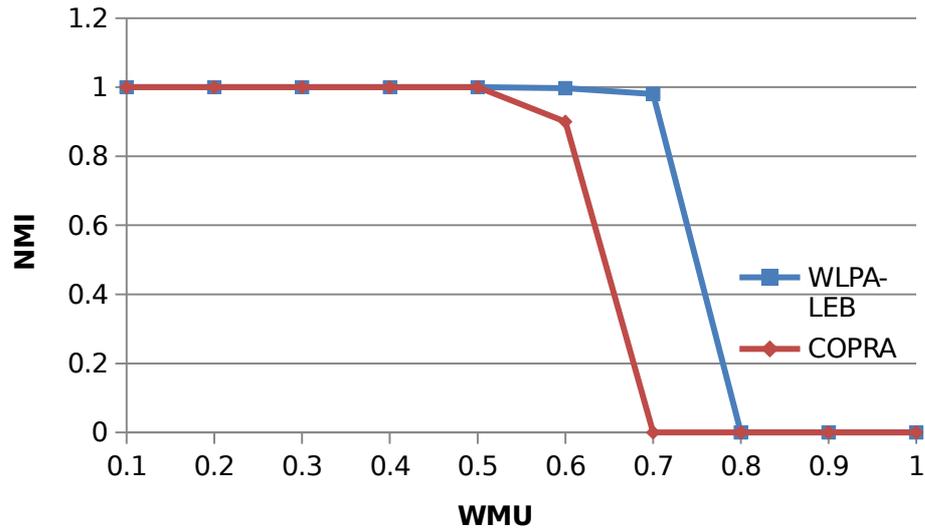

Figure 9, performance of WLPA and COPRA big synthesized weighted network with small communities (F). Results are compared through different value of *wmu*. When *wmu* exceeds 0.5, detected communities become unclear, yet WLPA-LEB could detects communities better than other algorithms.

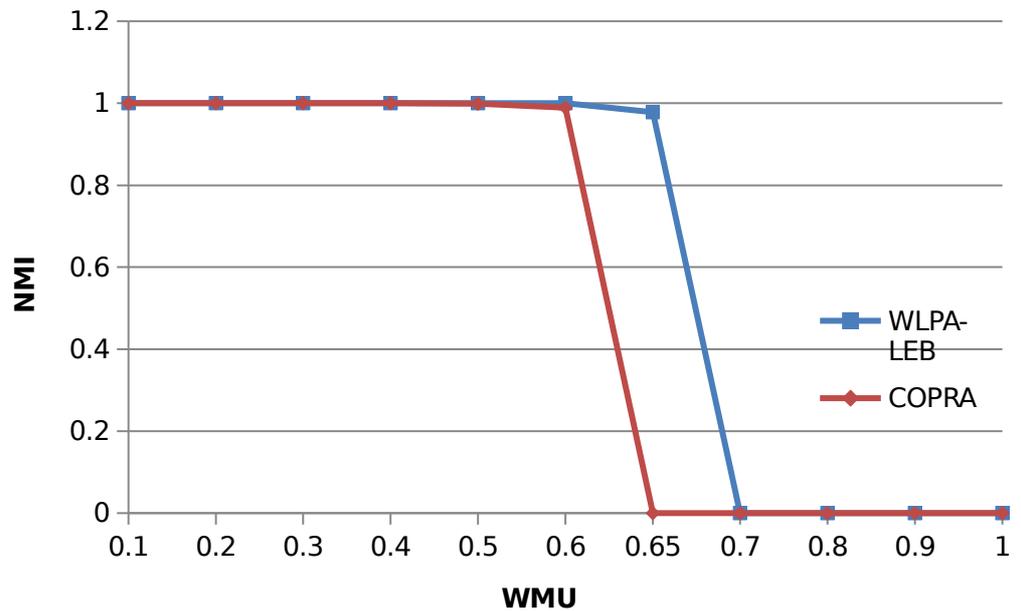

Figure 10, performance of WLPA and COPRA big synthesized weighted network with big communities (G). Results are compared through different value of *wmu*. When *wmu* exceeds 0.5, detected communities become unclear, yet WLPA-LEB could detects communities better than other algorithms.

In research [9], it has been claimed that LPA detects acceptable communities over 95% of networks with less than 5 iterations. We have evaluated the accuracy of WLPA-LEB with the maximum of 4 iterations for each answer which is represented in Table 5. As shown, WLPA-LEB could detect communities in less than 5 iterations same as traditional LPA.

Table 5, WLPA-LEB with 4 iterations could also detect communities in complex networks with acceptable accuracy.

| Network | Average modularity | | Best modularity | | Worst modularity | |
|---|---|---|---|---|---|---|
| | WLPA-LEB 4 iteration | WLPA-LEB | WLPA-LEB 4 iteration | WLPA-LEB | WLPA-LEB 4 iteration | WLPA-LEB |
| Karate | 0.3882 | **0.3906** | **0.4174** | 0.4155 | **0.1120** | 0.1120 |
| Dolphin | 0.5118 | **0.5152** | **0.5277** | 0.5264 | **0.4350** | 0.4350 |
| Football | 0.5959 | **0.5980** | **0.6045** | 0.6045 | **0.5568** | 0.5330 |
| Power | 0.7115 | **0.7844** | 0.7183 | **0.7794** | 0.7054 | **0.7887** |

During another experiment, we tried to examine the impact of the different depths of the local edge betweenness on the accuracy of detected communities. In Table 6, we have compared WLPA-LEB accuracy with different local edge-betweenness depths. As can be seen, deeper local edge betweenness does not have a significant effect on the accuracy of the algorithm, but for large-scale graphs, it is better to use local edge betweenness with depth 3 in order to increase the accuracy of the algorithm since communities may become big. However, as much as deeper local edge betweenness is calculated, the complexity increases; in addition, if we use traditional edge betweenness, the complexity of the entire algorithm will be $O(n.m)$.

Table 6, modularity of detected communities with different local edge-betweenness depths. Deeper local edge betweenness improves accuracy of detected communities but the improvement is not significant.

| Any depth | | depth-3 | | depth-2 | | Network |
|---|---|---|---|---|---|---|
| WLPA-LEB 4 | WLPA-LEB | WLPA-LEB 4 | WLPA-LEB | WLPA-LEB 4 | WLPA-LEB | |

| iteration | | iteration | | iteration | | |
|---|---|---|---|---|---|---|
| 0.4003 | **0.4012** | 0.3982 | 0.3991 | 0.3882 | 0.3906 | Karate |
| 0.5092 | 0.5094 | 0.5116 | 0.5141 | 0.5118 | **0.5152** | Dolphin |
| 0.5985 | 0.6002 | 0.5992 | **0.6016** | 0.5959 | 0.5980 | Football |
| 0.7295 | **0.8062** | 0.7180 | 0.7907 | 0.7115 | 0.7844 | Power |

**WLPA-LEB complexity**

In this part, we try to investigate the complexity of WLPA-LEB. As claimed, the algorithm has a linear time complexity in sparse graphs, and experiments have shown that the runtime is linearly related to the size of the problem. In order to practically compare the complexity of the proposed algorithm, runtime of WLPA-LEB and LPA is compared in a synthesized network with a moderate degree 15 and *mu* 0.4 which number of node ranges between 1000 and 1,000,000. We consider the average degree equal to 15 since the largest real graph, which is over 68 million, has an average degree 20. Figure 11 shows the output results, which increases linearly by increasing the number of vertices. A very interesting point in this experiment is the results obtained for graphs below 50,000 vertices in which the proposed algorithm finds answer faster while it is more complex than LPA. The reason is that, in small graphs, the cost of calculating local edge-betweenness is low, and WLPA-LEB could detect communities in lesser iterations by help of this metric; thus, the overall time decreases.

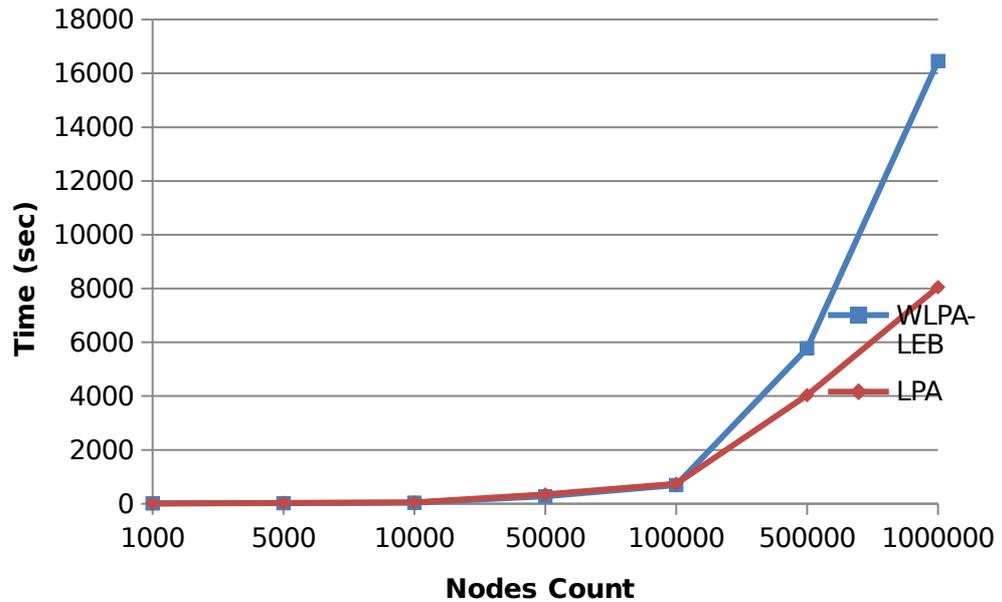

Figure 11, compare runtime of WLPA-LEB and traditional LPA which is $O(m)$. WLPA-LEB time complexity is as average degree higher than LPA, and average degree is usually constants in real-world networks.

In another experiment, we tried to examine the computation cost of local edge-betweenness with different depths. The network used in this experiment is a synthesized network with a moderate degree 15 and *mu* 0.4 which number of node ranges between 1000 and 1,000,000, and the results are summarized in Figure 12. The results indicate that calculating 2-depth local edge betweenness has a reasonable computational volume; even, 3-depth local edge-betweenness is measurable. Needless to say, calculating traditional edge betweenness is highly time-consuming.

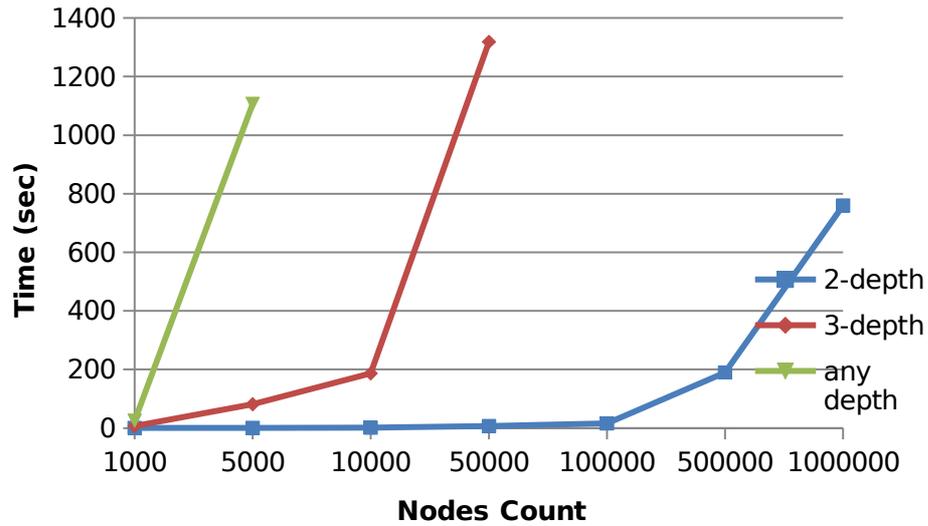

Figure 12, Compares the time required to calculate local edge-betweenness with different depths. In large graphs, the calculation of the edge betweenness with any depth is $O(n.m)$ and is not calculated for graphs above 50,000 vertices.

**Scalability**

In this part of the research, we intend to examine the scalability of the proposed algorithm in practice. Our experiments have also shown that the WLPA-LEB calculations are parallel and scalable, and as number of processor increases, runtime decreases relatively. To parallelize the algorithm, it should be possible to divide the algorithm into a number of microprocessors that have at least critical points. To do this, we consider the processing of each node, including taking label from neighbors and calculating local edge-betweenness as independent task. That is, each node tries to find most frequent label independently and in parallel. In the case of the calculation of local edge-betweenness, it is also possible to calculate the shortest path between the two vertices independently, which should only consider the critical conditions when updating the score of edges.

To parallelize this algorithm, we have used shared memory architecture and computed the effect of increasing the number of processors on runtime, as summarized in Figure 13. As can be seen, the speed of the algorithm increases as the number of processors increases. Generally, in shared memory mode, we do not suffer from message passing delays and limitations, so level of scalability is linear.

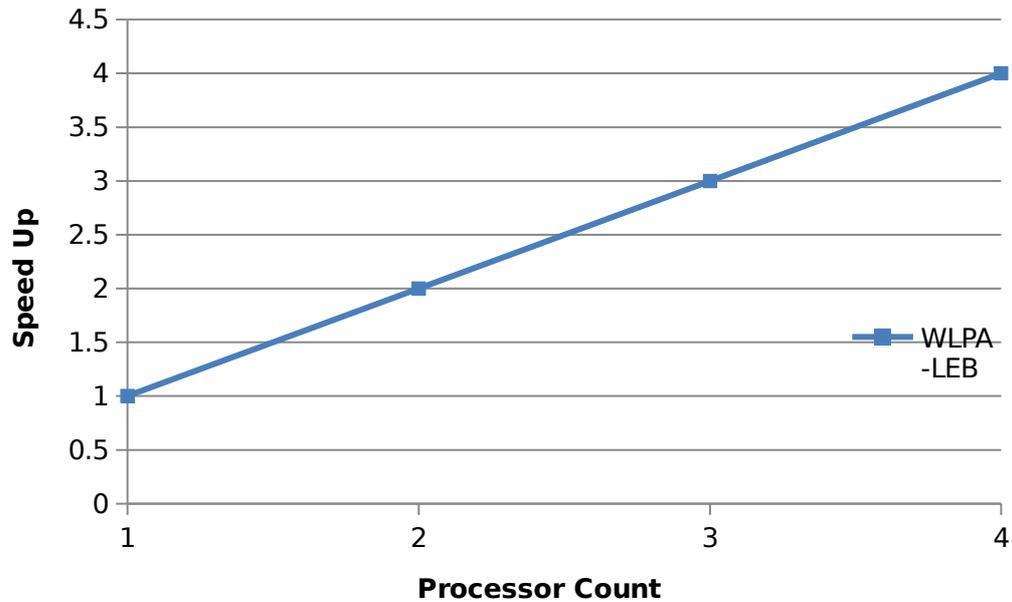

Figure 13, the effect of an increasing number of processors on runtime of WLPA-LEB in shared memory model. As the processors number increases, WLPA-LEB could detect communities faster.

## Conclusion

In this paper, we tried to describe community structures in complex networks and general techniques for detecting communities. Although many algorithms have been proposed, the main goal is increasing the accuracy of detected communities while the computation is not so time-consuming; hence, we proposed Weighted Label Propagation based on Local Edge Betweenness, abbreviated WLPA-LEB. The proposed algorithm is a combination of Girvan-Newman and LPA that improves the LPA randomization technique with the help of the Girvan-Newman edge betweenness, which enables it to efficiently identify communities in weighted networks. Experiments and practical comparisons with other community detection algorithms such as LPA, COPRA, Girvan-Newman, and LPAc all indicate that the proposed algorithm is capable of identifying communities with high precision. In the course of the evaluations, it was found that this algorithm detects communities with linear time complexity in sparse networks; also, the algorithm is scalable and can be implemented in parallel.

## Future Work

Community detection is still a hot topic in the complex networks field because its results are used in other areas, such as data analysis. Usually, social networks are dynamic that is the network properties change over time; for example, in social networks over time, the number of vertices (users) and the edges (communications) and the syntax of the relationship are changed because the users are active and dynamic, affecting the network. After a limited time period, the communities found on the old network are not correct in the current network. WLPA-LEB is not modified to detect communities in dynamic networks. Besides, social networks often are directed that means we must detect communities based on edges' direction, yet WLPA-LEB is not capable of identifying communities in directed networks. On the other hand, the proposed algorithm only recognizes non-overlapping communities, while overlapping communities have many applications, especially in the field of anomaly detection, which has not been considered in this research. Various algorithms are presented in this field, but the accuracy and speed of these algorithms can also be improved.